\documentclass[prb,amsmath,amssymb,letterpaper,twocolumn]{revtex4-1}

\usepackage{graphicx}
\usepackage{dcolumn}
\usepackage{bm}
\usepackage{epstopdf}
\usepackage{latexsym}
\usepackage{epsfig}
\usepackage{verbatim}
\usepackage{hyperref}
\usepackage{color}

\begin{document}

\title{Zeeman-limited Superconductivity in Crystalline Al Films}

\author{P.W. Adams}
\affiliation{Department of Physics and Astronomy, Louisiana State University, Baton Rouge, Louisiana 70803, USA}
\author{H. Nam and C.K. Shih}
\affiliation{Department of Physics, The University of Texas at Austin, Austin, TX 78712}
\author{G. Catelani}
\affiliation{Forschungszentrum J\"ulich, Peter Gr\"unberg Institut (PGI-2), 52425 J\"ulich, Germany}

\date{\today}

\begin{abstract}
 We report the evolution of the Zeeman-mediated superconducting phase diagram (PD) in ultra-thin crystalline Al films.  Parallel critical field measurements, down to 50 mK, were made across the superconducting tricritical point of films ranging in thickness from 7 ML to 30 ML.  The resulting phase boundaries were compared with the quasi-classical theory of a Zeeman-mediated transition between a homogeneous BCS condensate and a spin polarized Fermi liquid.  Films thicker than $\sim~20$ ML showed good agreement with theory, but thinner films exhibited an anomalous PD that cannot be reconciled within a homogeneous BCS framework.
 \end{abstract}


\maketitle


Tunable spin-imbalance offers a compelling probe of spin correlations, particularly in systems which have a macroscopic ground state that is incompatible with unequal spin populations. This subject has had a long history, but nevertheless, remains at the forefront of condensed matter and atomic physics. In condensed matter one of the most intensely studied examples is that of spin-singlet superconductors subjected to Zeeman and/or exchange fields. In the 1960's it was proposed that a Zeeman field could induce a spatially modulated order parameter in a spin singlet superconductor, known as the Ferrel-Fulde-Larkin-Ovchinnikov (FFLO) state \cite{FF,LO}. Over the last decade substantial thermodynamic evidence for its existence has emerged from studies of ultra-low impurity bulk superconductors such as the heavy fermion inter-metallic CeCoIn$_5$ \cite{Radovan,Kout} and the layered organic superconductors \cite{Beyer, Coniglio, Bergk}. For spintronics applications, the focus is on the interplay between superconductivity and ferromagnetism \cite{Linder}. For example, spin imbalance can be created in a superconductor by injecting spin-polarized currents from a ferromagnetic metal \cite{Quay}, or a ferromagnetic insulator can induce in the superconductor a large exchange field which can then be modulated by an applied magnetic field \cite{Xiong}. In cold atomic gases, an analog of FFLO has been proposed  \cite{Sheehy,Liao} whose behavior is affected by the effective dimensionality of the system. In this article we map out, as a function of temperature and film thickness, the Zeeman-limited superconducting phase diagram of crystalline Al films, which are effectively two-dimensional. The phase diagrams of films thinner than 20 monolayers have a structure that markedly differs from that expected for a homogeneous ground state. Our data add further evidence that these otherwise classical BCS superconductors evolve a non-trivial order parameter, that is neither homogeneous nor FFLO, when the Zeeman energy approaches the superconducting gap energy.

	The temperature dependence of the parallel (to the film surface) critical magnetic field was measured on epitaxial superconducting Al films, having thicknesses that varied between 7 ML (17 \AA) and 30 ML (72 \AA).  These thicknesses are much less than superconducting coherence length of the films $\xi\sim300$~\AA. In this limit, the orbital response to the field is suppressed, and a 1${\rm st}$-order transition to the normal state occurs when the Zeeman splitting is of the order of the superconducting gap $\Delta_0$ \cite{Fulde}.  The conventional picture is that this Zeeman-mediated transition, which is often referred to as the spin-paramagnetic transition, occurs between a homogenous BCS ground state and a polarized Fermi liquid normal state \cite{MT}.  The Zeeman critical field is expected to be near the Clogston-Chandrasekhar \cite{Clogston,Chandra} value $\mu_{\rm B}H_{cc}=\Delta_0/\sqrt{2}$, where $\Delta_0\approx 1.76k_{\rm B}T_c$ is the zero temperature gap, and $\mu_{\rm B}$ is the Bohr magneton.
	
\begin{figure}
\begin{flushleft}
\includegraphics[width=.44\textwidth]{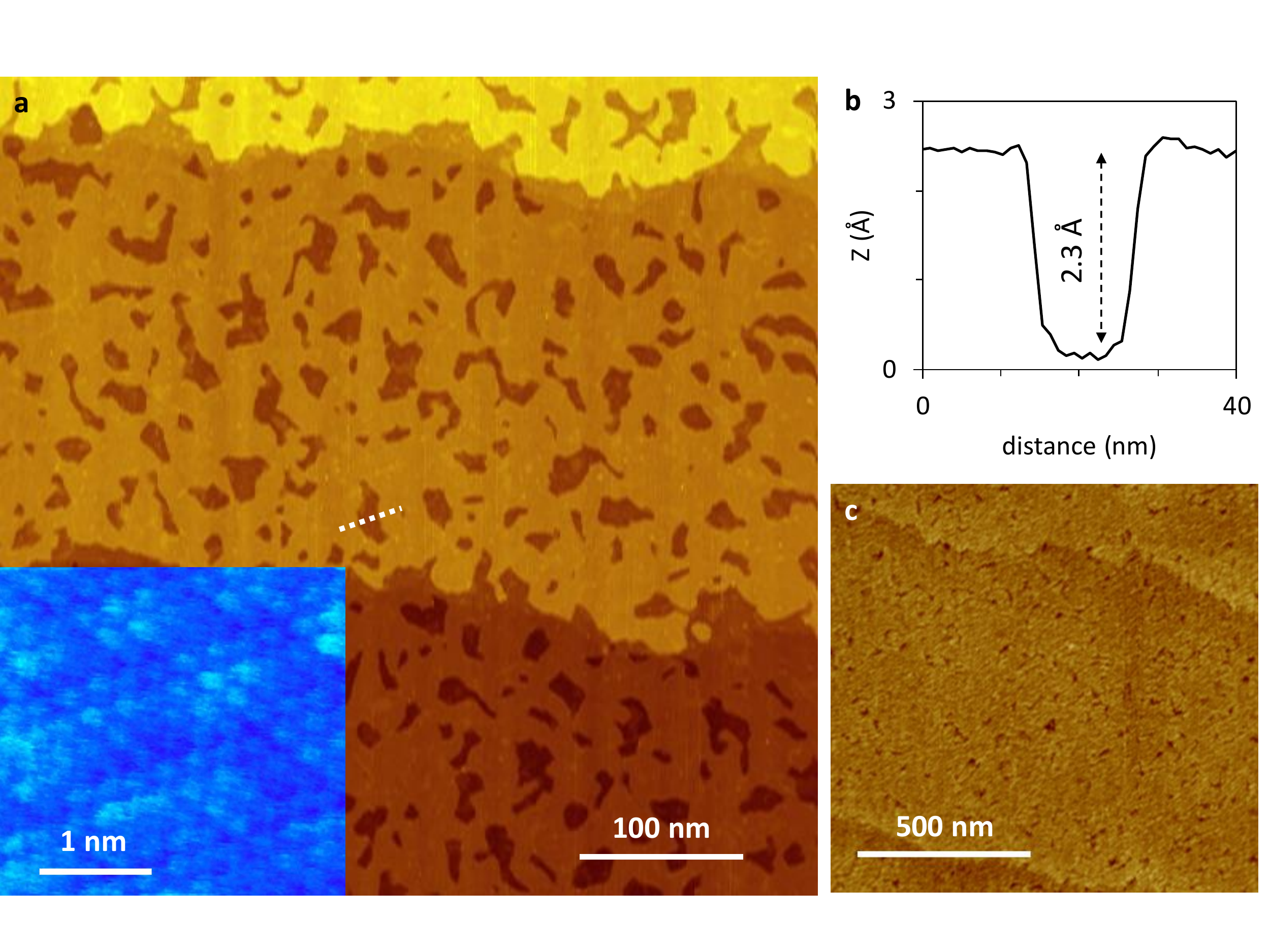}\end{flushleft}
\caption{\label{Fig1}{\bf a} In-situ STM image of a 10 ML-thick epitaxial Al film which shows atomically flat plateaus interspersed with 1 ML-deep pits. {\bf b} Profile trace along the white dash line which crosses over a pit.  {\bf c} AFM image of a 10 ML Al film capped by its native oxide. }
\end{figure}

	Epitaxial Al films \cite{Smith,Liu} were grown via a two-step method.  First, Al was deposited from a Knudsen cell at 0.5 \AA/min on a Si(111)-7x7 surface which was held below 100 K. After the low temperature deposition, the films were naturally annealed up to room temperature (RT). Shown in panel {\bf a} of Fig.\ \ref{Fig1} is an in-situ STM image of a 10 monolayer (ML) Al film, measured at 77 K, which shows an atomically flat surface interspersed with pits. A profile scan across a pit (see white dash line in Fig.\ \ref{Fig1}) reveals a depth of $\sim2.3$~\AA, corresponding to a 1-ML depth.  Panel {\bf b} of Fig.\ \ref{Fig1} clearly shows atomic ordering on (111)-surface. (See Appendix~\ref{app:epi} for further evidence of epitaxial growth.) For the ex-situ magnetotransport measurements, the epitaxial films were oxidized under an oxygen partial pressure of 1.6  $\mu$Torr for 10 min at RT.  This formed a AlOx capping layer. Panel {\bf c} of Fig.\ \ref{Fig1} shows an AFM image that was taken after the surface oxidation of an in-situ Al film. The silicon step edge and the pit features are clearly resolved, indicating that the capping layer formed without inducing significant damage to the underlying Al film.  We believe that the capping layer consumed approximately 3 - 4 ML of the exposed Al surface \cite{oxide}.  In all of the magnetotransport data presented below we conservatively estimate the metallic thickness of the films is to be 3 ML less than the as-grown thickness.  Therefore, the quoted film thicknesses in the phase diagrams represent an upper bound on the actual metallic thicknesses. Leads were attached to the films by first depositing Cr/Au contact pads via e-beam deposition and then soldering fine Pt wire to the contact pads with Wood's metal.  The magnetotransport measurements were performed on a dilution refrigerator equipped with a 9 T superconducting solenoid.  The films were aligned to parallel orientation with an in-situ mechanical rotator.
		
Previous magnetotransport measurements of the parallel critical field behavior of quench-condensed (QC) Al films revealed a hysteretic first-order critical field transition at temperatures below a tricritical point $T_{tri}\sim600$ mK \cite{Wu1,Wu2}.  Near the Zeeman critical field, QC films often exhibit non-equilibrium behavior such as stretched-exponential relaxations and avalanches.  Recent tunneling density of states measurements have shown that the avalanches represent irreversible collapses of macroscopic regions of superconductivity, and that they are not associated with magnetic flux jumps \cite{Presti}.  In addition to the unusual dynamics, $T_{tri}$ of QC Al films is typically a factor of two smaller than predicted by theory.   Because quench condensation produces a highly disordered, granular film morphology in Al \cite{Wu3}, one cannot easily assess which characteristics of Zeeman-limited superconductivity are attributable to disorder/morphological influences and which are a fundamental property of the condensate.  This issue is particularly pertinent to recent reports that disorder can stabilize a patchwork of FFLO-like superconducting puddles \cite{Presti,Loh}, despite the fact that it is generally agreed that the classic FFLO phase is suppressed in the presence of even modest disorder \cite{Radovan}.

In this study we have made detailed measurement of the Zeeman-limited superconducting phase diagram (PD) in epitaxial (ET) Al films of varying thickness and disorder.  As we show below, not only does epitaxial layer-by-layer growth give one unprecedented control of sample thickness for these types of studies, but for a given thickness, epitaxial films are substantially less disordered their QC counterparts.  This offers an unparalleled opportunity to study Zeeman-limited superconductivity in a system whose impurity density is far below what was previously attainable in metal films.

\begin{figure}
\begin{flushleft}
\includegraphics[width=.44\textwidth]{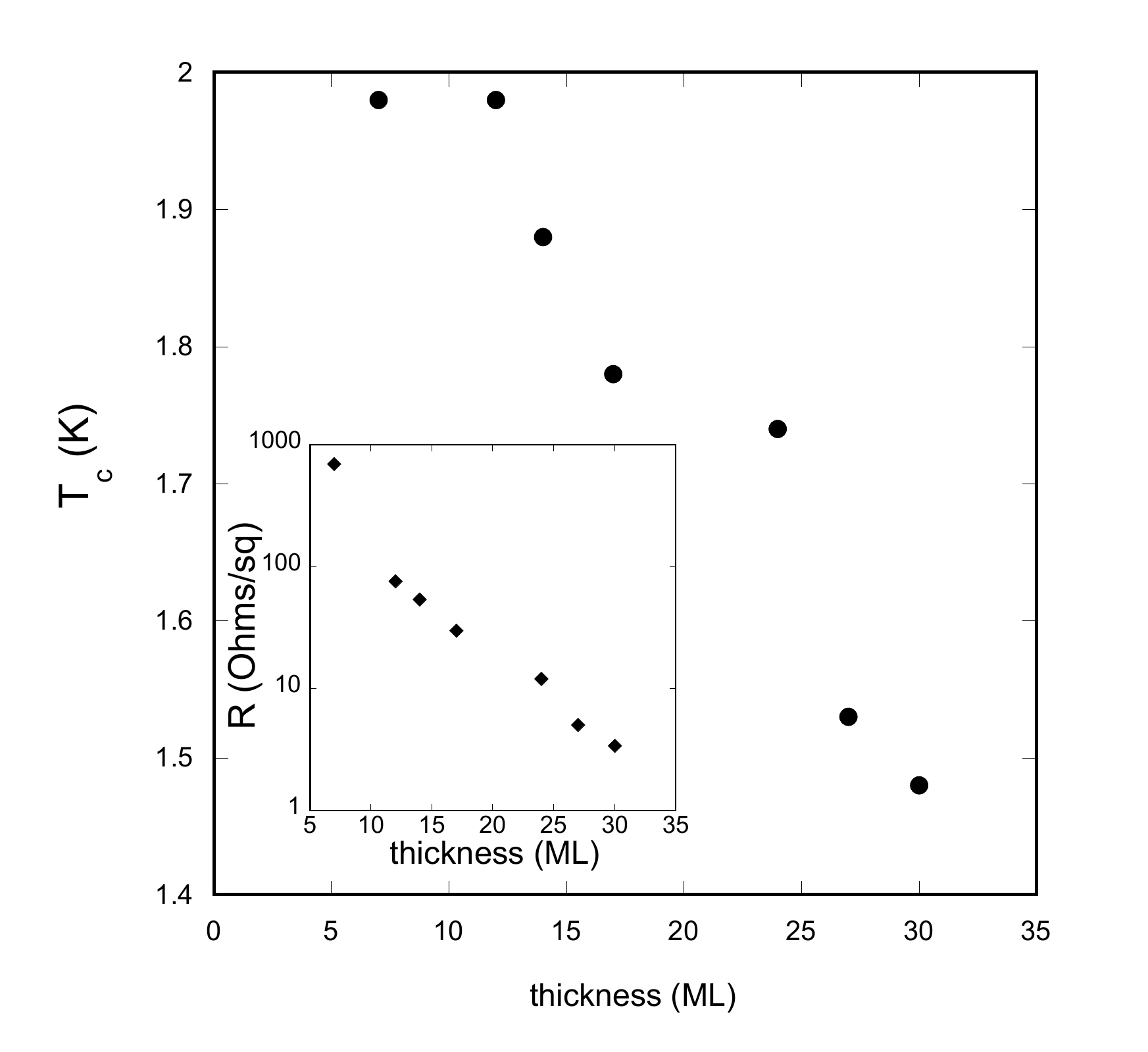}\end{flushleft}
\caption{\label{Fig2}The transition temperature of the epitaxial Al films used in theis study as a function of film thickness and sheet resistance.}
\end{figure}

Shown in Fig.\ \ref{Fig2} are the thickness and resistance dependencies of the transition temperature for a set of films ranging in thickness from 7 ML to 27 ML.  Note that the transition temperature rises rapidly with decreasing film thickness $t$ until it saturates at $\sim2$ K in films with $t\lesssim10$ ML.   This behavior cannot be attributed to the fact that the sheet resistance itself increases with decreasing $t$, see Fig.\ \ref{Fig2} inset.  Generally, amplitude fluctuations of the order parameter in homogeneously disordered superconducting films result in a reduction of $T_c$ as the films are made thinner and more resistive \cite{Valles}.

\begin{figure}
\begin{flushleft}
\includegraphics[width=.44\textwidth]{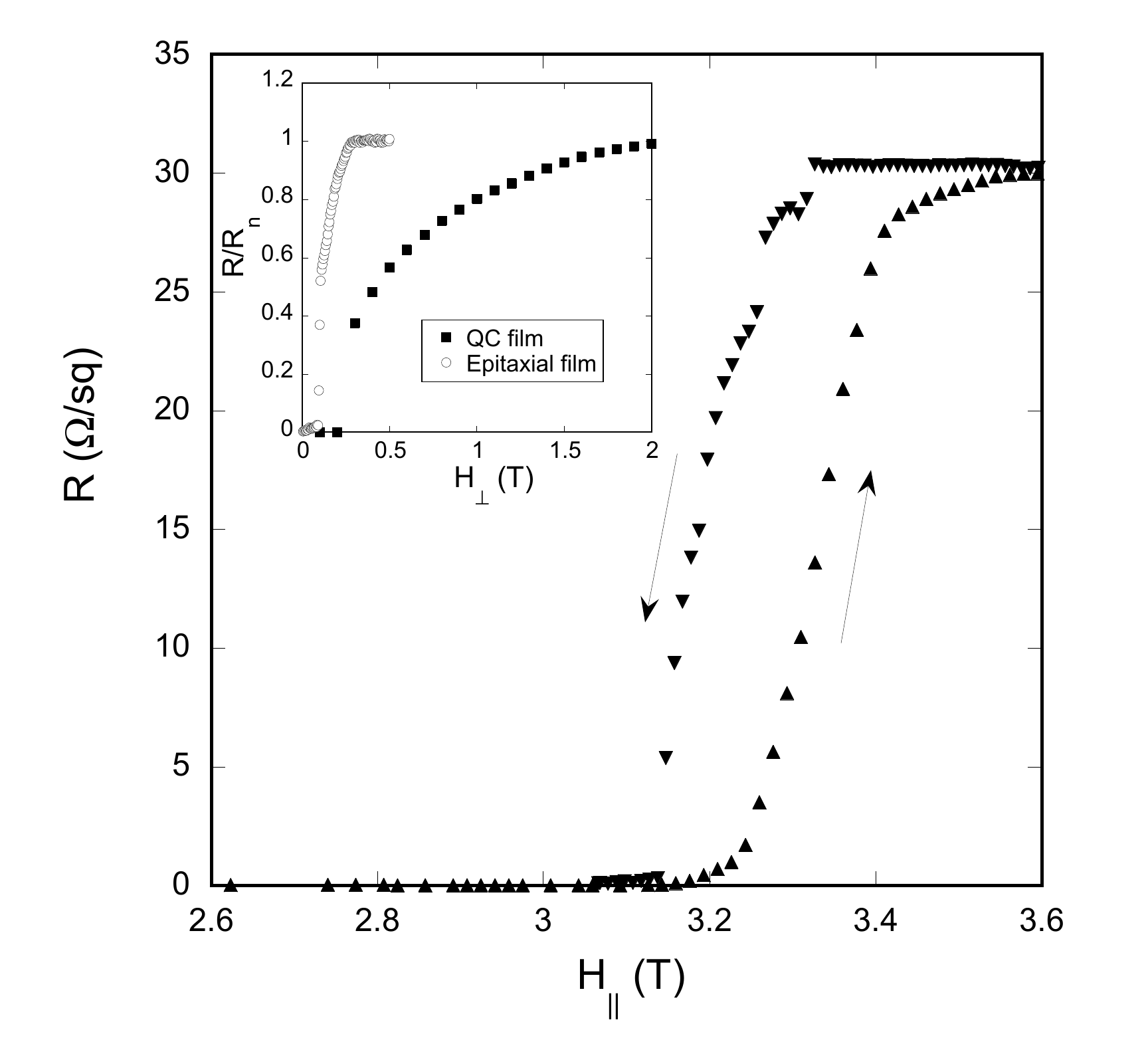}\end{flushleft}
\caption{\label{Fig3}Hysteretic parallel critical field transition of a 17 ML epitaxial Al film at 90 mK.  The arrows depict the magnetic field sweep direction.  Inset: Perpendicular critical field transition of a 12 ML epitaxial Al film and a comparably thick quench-condensed Al film.}
\end{figure}

The crystallinity of the ET films is reflected in the fact that their sheet resistances are a factor of 2 - 3 times {\it lower} than comparably thick QC films.   The differing disorder levels between these two types of films is also evident in their respective perpendicular critical field, $H_{c2}$, behavior.  For comparison, we produced a QC Al film which had the same as-deposited thickness  (48 \AA) as the 12 ML ET sample used in this study.  We assume that the two samples developed oxide layers of similar thickness and that the mean-free-path $l_o$ of each was much less than their respective coherence lengths.  In this ``dirty limit'' $H_{c2}=\dfrac{\Phi_o}{2\pi\xi_ol_o}$, where $\Phi_o$ is the flux quantum,  and $\xi_o\sim1600$ \AA~ is the BCS coherence length of bulk Al \cite{Tinkham}.  The QC film had a transition temperature $T_c=2.4$ K, normal state sheet resistance $R=84~\Omega$, and $H_{c2}=2.0$ T as measured at $T = 0.5$ K.  In contrast, the 12 ML ET film had a $T_c=2.0$ K, $R=30~\Omega$. and $H_{c2}=0.28$ T, see inset of Fig.\ \ref{Fig3}.   From these data we can extract the respective ratios of the Pippard coherence length and the mfp for the two types of films: $\xi^{ET}_o/\xi^{QC}_o\sim 3$ and $l^{ET}_o/l^{QC}_o\sim 6$.

 Figure\ \ref{Fig3} shows an example of a typical resistive parallel critical field transition of a 17 ML Al film taken at 90 mK.  The hysteresis is indicative of the 1$^{\rm st}$-order transition, which was observed in all of the films studied except the 30 ML sample.   In contrast to QC films, we found no evidence of avalanches in the critical field traces of any of the samples in this study.  By measuring the hysteresis loops as a function of temperature and thickness one can map out the entire Zeeman-limited PD.  We define the critical field at the midpoint of the transition and then plot the temperature dependence of the up-sweep (superheating) and down-sweep (supercooling) critical fields. Although the midpoint criteria is arbitrary, the overall structure of the resulting phase diagrams does not vary significantly when one uses a different criteria for $H_c$ such as when the resistance reaches 10\% of the normal state resistance or when it reaches zero (see Appendix~\ref{app:width}).   In addition to the finite width of the critical field transitions, the analysis is complicated by the fact that, in the hysteretic region, the films are in a metastable state and therefore exhibit some temporal relaxation.  Because of this the width of the hysteresis loops is a weak function of the magnetic field sweep rate.  Slower sweep rates produce slightly narrower hysteresis loops.  However, the salient features of the phase diagrams remain unchanged when the sweep rate is varied.

\begin{figure}
\begin{flushleft}
\includegraphics[width=.44\textwidth]{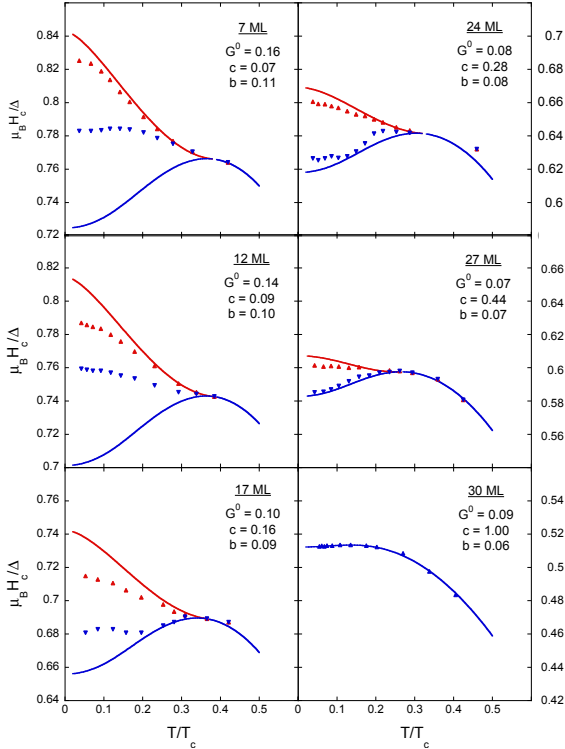}\end{flushleft}
\caption{\label{Fig4}Zeeman-limited phase diagrams of epitaxial Al films of varying thickness.  The symbols represent the superheating (upward triangles) and supercooling (downward triangles) critical fields as a function of reduced temperature.  Note that the abscissa field scale is the same for each panel.  The superconducting gap was determined from the transition temperature via the BCS relation $\Delta_0=1.74kT_c$.  The lines are the theoretic phase boundaries as obtained from QCST by varying $G^0$, $b$, and $c$.  The best fit values of these parameters are listed in the panel legends. The tricritical point is defined by the temperature at which the parallel critical field transition becomes hysteretic.}
\end{figure}

   Figure \ref{Fig4} shows the resulting PD of six samples that range in thickness from a few monolayers to 30 monolayers.  The abscissa scale of each panel is the same.  The triangular symbols are the measured reduced critical fields, which are normalized by the superconducting gap $\Delta=1.76 k_{\rm B}T_c$.  The upward triangles (red symbols) represent the superheating phase boundary and the downward triangles (blue) the supercooling boundary.  The solid lines are fits to weak-coupling superconductivity theory, which assumes that the transition occurs between a homogeneous BCS ground state and a polarized Fermi liquid.

The superconducting properties of thin films in the presence of high Zeeman field are influenced by (1) Fermi-liquid effects which renormalize the spin susceptibility, (2) spin-orbit scattering which inhibits spin polarization, and (3) sample thickness, which determines the relative importance of the orbital response to the magnetic field.   The quasi-classical theory of weak coupling superconductivity \cite{Eilenberger,Larkin} (QCTS), as applied to the Zeeman-limited superconductivity \cite{Alexander,Suzuki,Catelani}, captures these effects via the corresponding dimensionless parameters \cite{Maki}: the anti-symmetric Fermi-liquid $G^0$ , the spin-orbit $b=\hbar/(3\tau_{so} \Delta_0)$, where $\tau_{so}$ is the spin-orbit scattering time, and the orbital pair-breaking $c\propto D t^2$, where $D$ is the electron diffusivity and $t$ is the film thickness.  $G^0$ is a measure of the renormalization of the spin susceptibility of an interacting Fermi gas.  It is related to the ratio of the spin susceptibility density of states $N_\chi$ to the specific heat density of states \cite{Baym} $N_\gamma$ by $G^0=N_\gamma/N_\chi-1$.

\begin{figure}
\begin{flushleft}
\includegraphics[width=.44\textwidth]{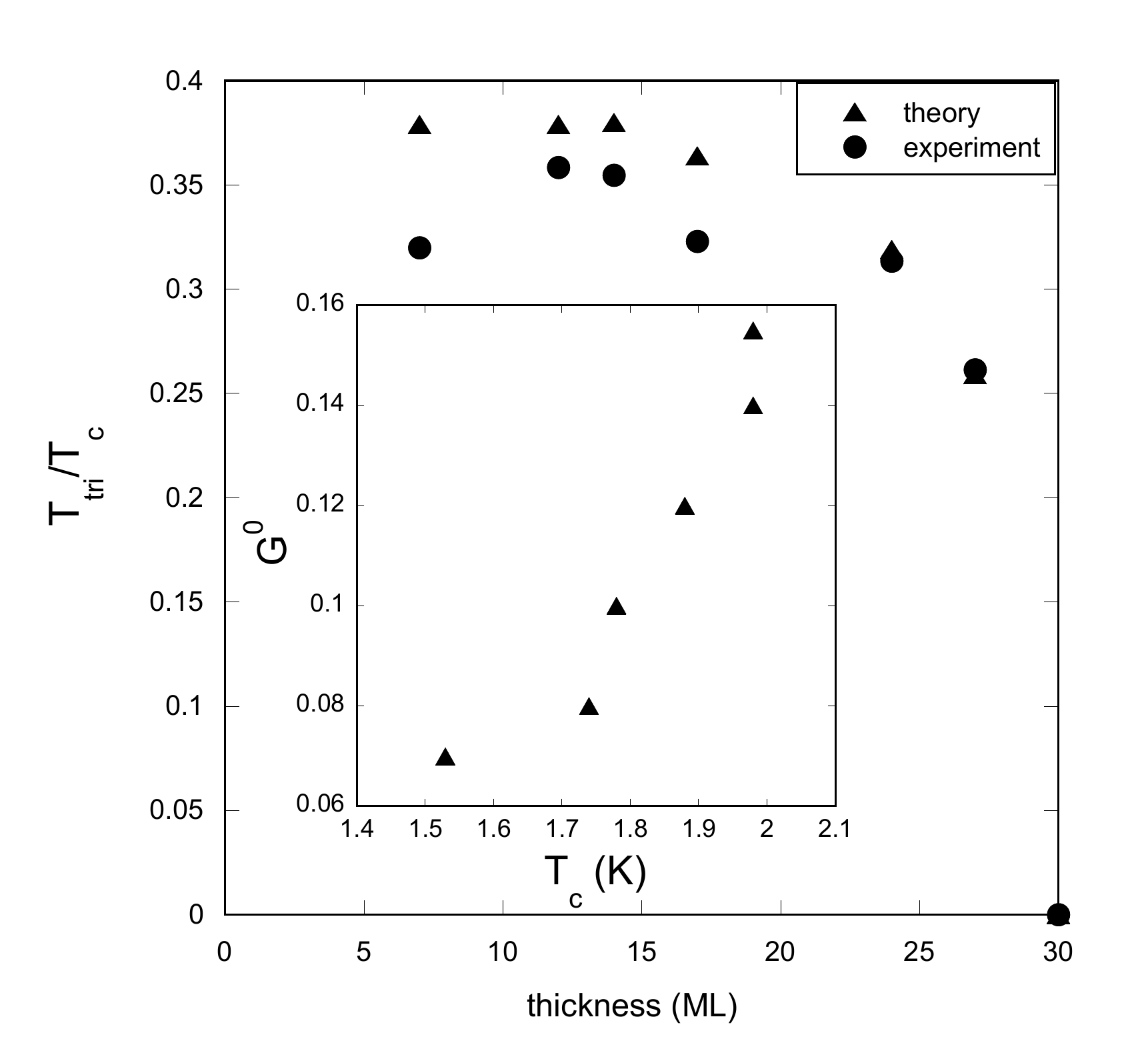}\end{flushleft}
\caption{\label{Fig5}Reduced tricritical point temperature as a function of film thickness.  The triangles were obtained from the QCST fits and the circles from the critical field measurements.  Inset: The anti-symmetric Fermi liquid parameter obtained from the QCST fits as a function $T_c$.}
\end{figure}

The QCST traces in Fig.~\ref{Fig4} where obtained by varying $G^0$, $b$, and $c$ in order to get the best correspondence to the measured phase diagram.  Details of this procedure are provided in the Appendix~\ref{app:theory}.  Following the evolution of the PD's in Fig.\ \ref{Fig4} from the thickest films to the thinnest, we first note that the critical field transition in the 30 ML sample remains 2$^{\rm nd}$-order down to the lowest temperatures measured ($\sim 70$ mK).  Also note that there is an excellent agreement between theory and the measured phase boundaries.  Furthermore, the extracted values of $G^0$, $b$, and $c$ are consistent with results from studies of relatively thick QC Al films \cite{Catelani}. Interestingly, the spin-orbit parameter $b$ increases with decreasing thickness, see Appendix~\ref{app:so}.  This suggests that a small but measurable spin-orbit scattering rate is associated with the Si-Al interface \cite{Adams}.  Therefore, as the film thickness is lowered the interface contribution to $b$ becomes more significant.

The antisymmetric Fermi liquid parameter, $G^0$, which accounts for the spin-triplet interaction channel, also increases with decreasing film thickness. The origin of this thickness dependence is unknown, but $G^0$ does appear to track the thickness dependence of $T_c$, see inset of Fig.\ \ref{Fig5}.      This implies that the underlying mechanism that gives rise to the enhancement of the spin-singlet interaction channel, which is reflected in $T_c$ also affects the spin-triplet channel and, consequently, the normal state spin susceptibility.

As can be seen in the 27 ML panel of Fig.\ \ref{Fig4}, decreasing the thickness by only 3 ML reduces the orbital depairing rate enough to open a 1$^{\rm st}$-order transition below a tricritical point $T_{\rm tri}\sim 380$ mK.  Both the tricritical point and the temperature dependence of the hysteresis width $\Delta H_c(T)$ are well accounted for by the theory.  But as the film thickness is decreased further, the measured PD's begin to deviate more and more from the theoretical curves.  Although QCTS can account for $T_{\rm tri}$ across the entire range of thicknesses, see Fig.\ \ref{Fig5}, the measured hysteresis magnitudes are much smaller than expected in the thinner samples.   Even more striking, the slopes of the down-sweep branches of the 12 and 7 ML PD's are either flat or slightly negative whereas the slopes of the theory traces are robustly positive.  Since the down-sweep critical fields represent the transition from the normal state to the superconducting state, the data in the 12 ML and 17 ML panels indicate that the superconducting phase nucleates well before theory would predict.  This behavior is somewhat counterintuitive. It suggests that the non-equilibrium normal state is more fragile than the corresponding superconducting state.

One possibility is that the metastable normal state is simply more susceptible to environmental fluctuations than the superconducting phase which prevents the system from reaching the theoretical supercooling phase boundary.  Another possibility is that quantum fluctuations about an intermediate inhomogeneous phase compromise the free energy barrier associated with the 1$^{\rm st}$-order transition to the superconducting phase.  Indeed, the relative asymmetry of the superheating and supercooling phase boundaries, as compared to the corresponding theory traces, is reminiscent of the asymmetric avalanche behavior observed near the Zeeman critical field of QC Al films \cite{Presti}.  Specifically, highly disordered QC Al films often exhibit avalanche-like jumps in the superheating branch of the hysteresis loop but only very rarely are avalanches observed on the supercooling branch.  The absence of supercooling avalanches is consistent with the fact that the supercooling branches of the 7, 12, and 17 ML Al films never approach the theoretical limit of metastability.

It is somewhat surprising that the QCST description of the Zeeman-limited PD breaks down in the regime where the orbital pair-breaking contributions are completely negligible.  If the films were, in fact, free of disorder, this is precisely the regime where one would expect the FFLO phase to emerge.  Interestingly, recent Hubbard model calculations have shown that near the Zeeman critical field a vestige of an FFLO-like phase is stabilized by a finite impurity density \cite{Loh,Yang}.  This disordered-LO phase is associate with local modulations of the pairing amplitude which, of course, should exhibit some manifestation in the structure of the PD.   We speculate that this inhomogeneous phase is preempting the expected supercooling critical field.  Extending the present work to include spin-resolved tunneling probes of the Zeeman-limited condensate may help confirm this possibility.

\acknowledgments

  The magntotransport measurements were performed by P.W.A. with the support of the U.S. Department of Energy, Office of Science, Basic Energy Sciences, under Award No.\ DE-FG02-07ER46420.  Film fabrication and characterization was performed by H.N. and C.K.S. with support from grants ONR-N00014-14-1-0330 and NSF-DMR-1506678.  The theoretical analysis was carried out by G.C. with partial support by the EU under REA Grant Agreement No. CIG-618258.

\appendix

\section{Evidence of Epitaxial Growth}
\label{app:epi}

The lattice constant ratio between Al and Si is $1.34$, very close to a 4:3 ratio, thus enabling the possibility of a co-incidental lattice match. This epitaxial growth condition is confirmed with three independent experimental techniques: Direct observation of atomic arrangement using in-situ STM, in-situ RHEED measurements of the surface structure during the growth and ex-situ XRD measurements after surface is passivated.  STM measurements reveal a uniform and smooth Al film with underlying steps of the Si(111) substrate in Fig.~\ref{Fig1}(a). Shown in the inset of Fig.~\ref{Fig1}(a) is the atomically ordered (111) surface of in-situ Al film with the lattice constant of $\sim$ 2.8~\AA. The measured lattice constant is consistent with the value of 2.86~\AA\, for bulk Al(111) surface, providing strong evidence that our in-situ Al film is grown epitaxially.

\begin{figure}
\begin{flushleft}
\includegraphics[width=.48\textwidth]{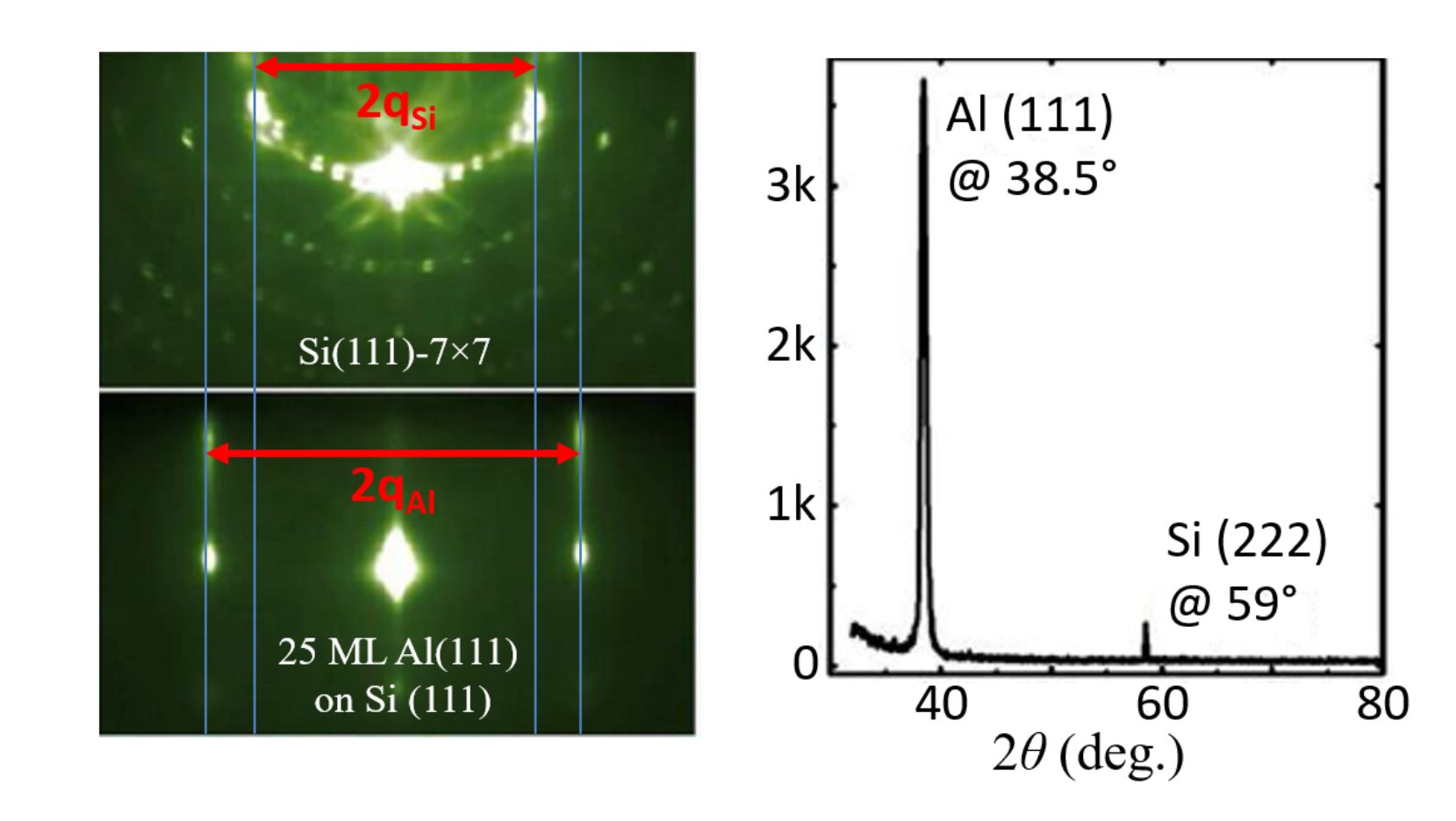}\end{flushleft}
\caption{\label{Fig8}(a) RHEED patterns of a Si(111)-$7\times7$ surface (upper) and an in-situ 25 ML Al film deposited on Si(111), respectively. Note that $q_{\rm Al}$/$q_{\rm Si}$ $\approx\frac{4}{3}$. (b) XRD pattern of a 100 ML Al film on Si (111). Although not apparent here, the Si(111) diffraction peak is located at $2\theta = 28.5^\circ$.}
\end{figure}

Shown in Fig.~\ref{Fig8}(a) are RHEED patterns for the Si(111)-$7\times7$ surface and the subsequently grown, 25 ML (5.85 nm) Al(111) film, respectively. The electron beam projection direction on the surface is in parallel to the Si [$11\bar{2} $] direction. The closely spaced diffraction spots in the upper panel of Fig.~\ref{Fig8}(a) represents $7\times7$ and the brightest spots corresponds to the $1\times1$ diffraction spots. In the case of Al(111), there is surface reconstruction, and the spacing between the $1\times1$ spot is about $4/3$ times that between the Si(111)-$1\times1$ diffraction spots. This confirms the epitaxial relationship between Al(111) and Si(111) with a 3:4 ratio in lattice constant.
More precise determination of lattice constant is carried out using XRD as shown in Fig.~\ref{Fig8}(b). Here the Al(111) diffraction peak is observed at $2\theta = 38.5^\circ$. The Si(111) diffraction peak (not shown here) is observed at $2\theta = 28.5^\circ$. With these two experimental values of diffraction angles, indeed one can deduce a ratio in lattice constant between the Si(111) substrate and the Al(111) thin films to be $1.34$. It is also interesting to observe the Si(222) diffraction peak. For a bulk Si, the (222) peak is forbidden by the  diamond crystal structure. However, the presence of the Si/Al interface enables a weak Si(222) diffraction peak due to the breaking of translational symmetry. (For larger thicknesses of Al ($> 150$ nm), this peak is no longer observable).

\section{Effects of Finite Transition Width}
\label{app:width}

\begin{figure}
\begin{flushleft}
\includegraphics[width=.47\textwidth]{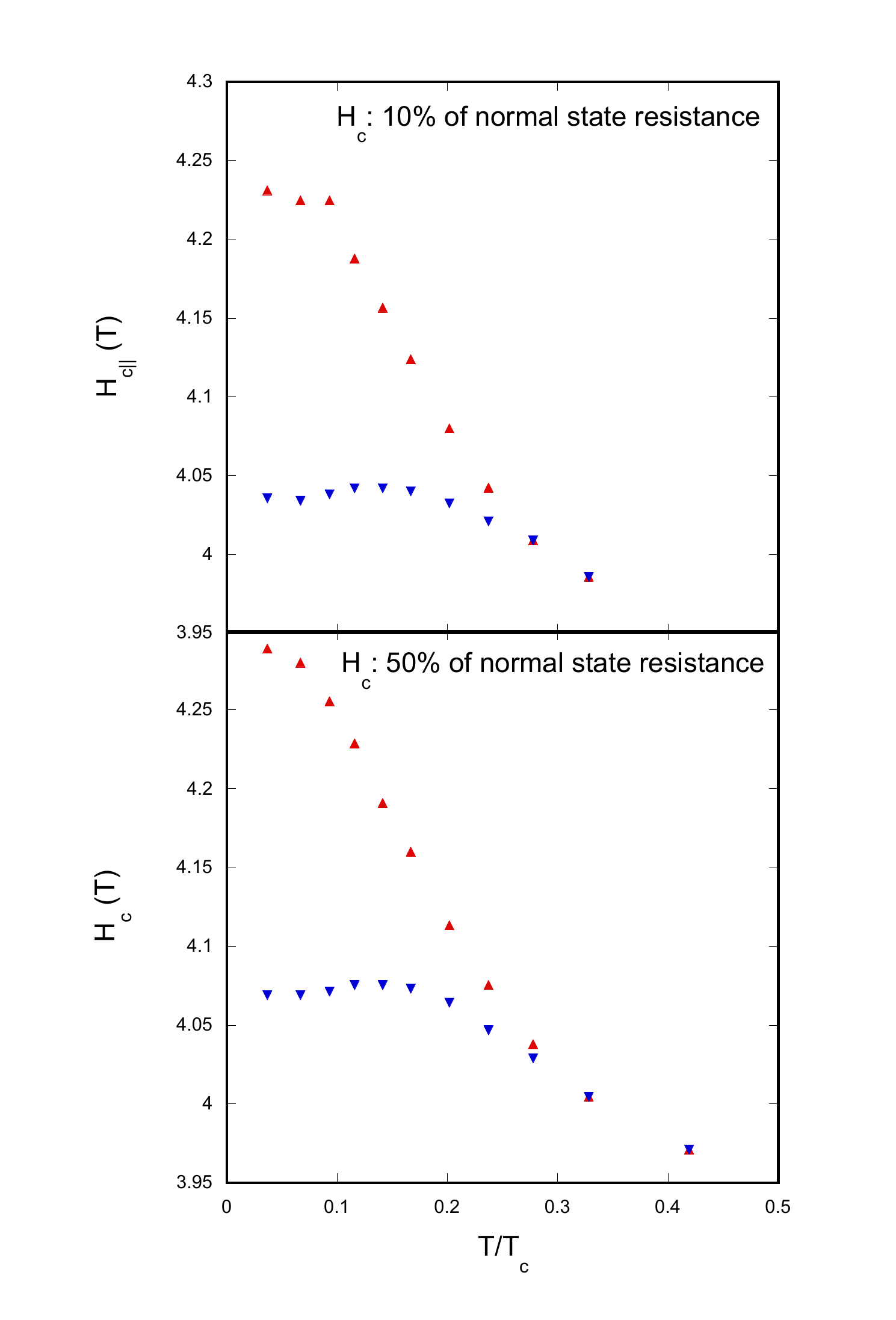}\end{flushleft}
\caption{\label{Fig7}Comparison of the phase diagrams of the 12 ML Al film that are obtained when the critical field is defined by the midpoint of the transition (lower panel) and by the foot of the transition (upper panel).}
\end{figure}

The finite width of the superconducting transitions introduces an ambiguity into the definition of the critical field and critical temperature.  Typically one defines the transition by its midpoint, but other criteria can also be used.  For instance, one can define the transition as the temperature or field at which the resistance reaches 10\% of its normal state value or 90\% (onset) of its normal state value or when the resistance falls to zero.  Of course, there is always a concern that the phase diagram depends on which definition of the critical field was used.  As can be seen in Fig.~\ref{Fig7}, using a 50\% or 10\% criterion for $H_c$ does not alter the overall structure of the phase diagrams in this study.  Therefore, the discrepancy with the quansi-classical BCS theory is not an artifact of the finite transition width.

\section{Theoretical Analysis}
\label{app:theory}

The thermodynamic and transport properties of thin superconducting films in the presence of a high parallel magnetic field are influenced by various material parameters. These include the film thickness $t$, which determines the relative importance of Zeeman splitting and the orbital effects of the magnetic field, the spin-orbit scattering time $\tau_{so}$, and Fermi-liquid effects that renormalize the spin susceptibility. For low-$T_c$ superconducting films, a quasiclassical generalization of BCS theory (QCST) incorporating these effects was developed in Ref.~\onlinecite{Alexander}; for numerical calculations, we find the reformulation presented in Ref.~\onlinecite{Suzuki} easier to implement. The three dimensionless parameters capturing the effects mentioned above are the antisymmetric Fermi-liquid parameter $G^0$, the spin-orbit parameter $b=\hbar/3\tau_{so} \Delta_0$ (with $\Delta_0$ the zero-temperature gap), and the orbital pair-breaking parameter $c$ [\onlinecite{Maki}]:
\begin{equation}
c = \frac{e^2 D t^2 \Delta_0}{6 \hbar \mu_B^2} \, g\left(\frac{\pi \ell_o}{t}\right)
\end{equation}
with $D$ the diffusion coefficient, $e$ the electron charge, $\mu_B$ the Bohr magneton, and $\ell_o$ the mean free path. This expression is valid for any ratio of mean free path to thickness, with the function $g$ defined as
\begin{equation}
g(x) = \frac{3}{2x^3}\left[\left(1+x^2\right) \arctan x - x\right]
\end{equation}

Given the above parameters, the task is to solve the so-called Usadel equations for the semiclassical
Green's functions, together with the self-consistent
equations for the order parameter $\Delta$ and the internal magnetic
field $H_i$; the latter takes into account the Fermi-liquid renormalization of the spin susceptibility, which in the normal state leads to $H_i = H_a/(1+G^0)$ with $H_a$ the applied field. For the calculation of the supercooling field and of the critical field above the tricritical temperature -- fields at which the order parameter vanishes and which we denote with $H_{c2}$, -- the solution of these equations can be found in analytical form, and an equation determining $H_{c2}$ as a function of temperature $T$ can be obtained; see Refs.~\onlinecite{Alexander,Suzuki}. This equation is then easily solved numerically.
In contrast, the calculation of the superheating field $H_{sh}$ requires a fully numerical solution of the Usadel and self-consistent equations. To find $H_{sh}$ we exploit the fact that for fields between $H_{c2}$ and $H_{sh}$ the self-consistent equation for the order parameter has two solutions, one stable and the other unstable. At the superheating field these two solutions collapse, implying that the derivative of the self-consistent equation with respect of the order parameter vanishes at that field; therefore, imposing this additional condition makes it possible to uniquely determine $H_{sh}$.

To compare theory and experiment, we proceed as follows: for the thickest film, we find a good fit to the experimental data by fixing the spin-orbit parameter $b=0.06$ to be similar to previously measured values and letting $c$ and $G^0$ vary freely. This approach gives the best fit to the low-temperature data, but cannot be applied to the thinner films which display hysteresis. For these films, we find the parameters that best agree with the experimental data subject to the following constraints: the orbital pair-breaking parameter $c$ should decrease with decreasing thickness; the calculated tricritical point should be as close as possible to the measured one, while at the same time the calculated superheating curve should always give fields higher than the measured $H_{sh}$, and the calculated supercooling curve should always be below the measured supercooling fields. This accounts for the fact that the calculated curves represent limits of stability for metastable states, so phase transitions should take place between these limits but not outside them.

\begin{figure}
\begin{flushleft}
\includegraphics[width=.44\textwidth]{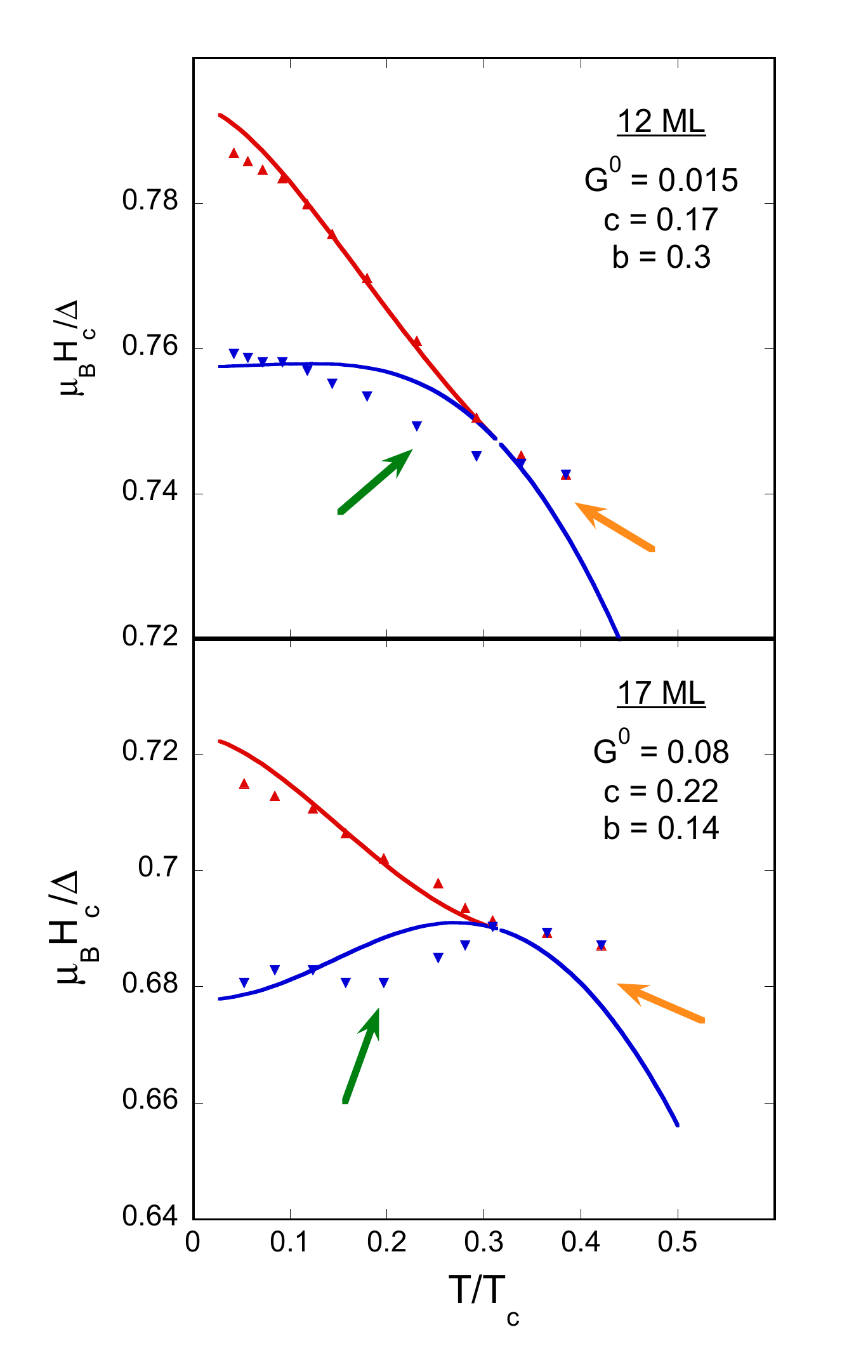}\end{flushleft}
\caption{\label{Fig9}Phase diagrams for the 12 ML and 17 ML films along with the corresponding fits.  The fits were obtained using only the data points below the tricritical point.  The arrows indicate regions of discrepancy between QCST and measurement. }
\end{figure}

In order to more fully explore the fitting parameter space we also performed fits in which we only used the data points below the tricitical point.  This optimization strategy places a much larger emphasis on the magnitude of the hysteresis.   As can be seen in Fig.~\ref{Fig9}, with this fitting strategy the theory accurately predicts the superheating branch and the overall hysteresis width, but it crosses a portion of the supercooling branch (see green arrows) and badly underestimates the higher temperature, $2^{\rm nd}$-order, critical fields (see orange arrows).   Since the theory traces represent the limits of metastability, it is unphysical for any portion of the supercooling branch to lie below the theory trace.  Furthermore the values of the fitting parameters obtained with this optimization scheme are not in good agreement with values obtained from previous studies.  The values of $G^0$ are a factor of 2 - 10 too small, and the spin-orbit parameter is a factor of 2 too large.  This analysis suggests that the discrepancy between theory and data cannot be resolved by a more judicial choice of fitting parameters.

\section{Spin-orbit Scattering Parameter}
\label{app:so}

\begin{figure}[h]
\begin{flushleft}
\includegraphics[width=.44\textwidth]{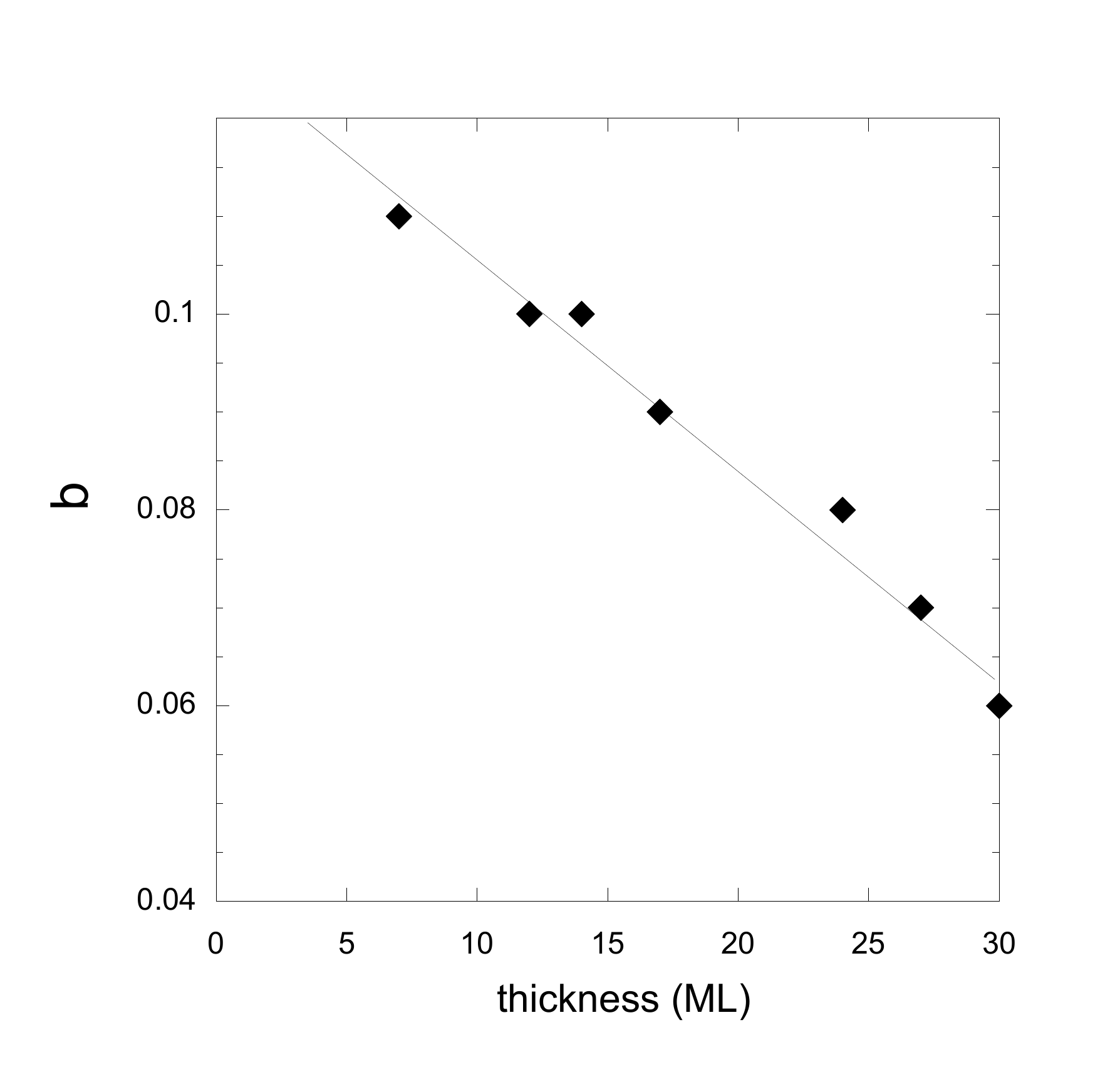}\end{flushleft}
\caption{\label{Fig6}Spin-orbit scattering parameter as a function of film thickness.}
\end{figure}

Shown in Fig.~\ref{Fig6} is the spin-orbit scattering parameter, as obtained from the phase diagram fits, as a function of film thickness.  Note that $b$ appears to vary linearly with thickness.  These data can be compared with those obtained by coating low atomic mass superconducting film with high atomic mass nobel metals.

Specifically, as reported in Ref.~\onlinecite{spinorbit} the spin-orbit coupling parameter in superconducting Be films of varying thickness was measured before and after coating the films with 0.5 nm layer of Au.  The Au layer induced a large spin-orbit scattering rate in the superconductor, which completely suppressed the tricritical point and corresponding hysteresis in the critical field transition.  In addition, the induced spin-orbit scattering greatly increased the parallel critical field of the films.   In contrast to our observations, the spin-orbit scattering parameter in the Be/Au bilayers varied as $b\sim\Delta_0/t^2$.  This discrepancy may be due to the fact that the epitaxial Al films are in the limit of $b\ll 1$ but, the Be/Au system was in the strong spin-orbit limit $b\ge10$.

\end{document}